# Temporal optical besselon waves

**Anastasiia Sheveleva [1], Ugo Andral [1], Bertrand Kibler [1],**
**Sonia Boscolo [2], and Christophe Finot [1,*]**

[1] *Laboratoire Interdisciplinaire Carnot de Bourgogne, UMR 6303 CNRS-Université de Bourgogne-Franche-Comté, 9 avenue Alain Savary, BP 47870, 21078 Dijon Cedex, France*

[2] *Aston Institute of Photonic Technologies, School of Engineering and Applied Science, Aston University, Birmingham B4 7ET, United Kingdom*

[*] *Corresponding author:*

E-mail address: christophe.finot@u-bourgogne.fr

Tel.: +33 3 80395926

**Abstract:** We analyse the temporal properties of the optical pulse wave that is obtained by applying a set of spectral $\pi/2$ phase shifts to continuous-wave light that is phase-modulated by a temporal sinusoidal wave. We develop an analytical model to describe this new optical waveform that we name 'besselon'. We also discuss the reduction of sidelobes in the wave intensity profile by means of an additional spectral $\pi$ phase shift, and show that the resulting pulses can be efficiently time-interleaved. The various predicted properties of the besselon are confirmed by experiments demonstrating the generation of low-duty cycle, high-quality pulses at repetition rates up to 28 GHz.

# I. Introduction

Ultrafast optics has provided extremely efficient means to generate various pulse waveforms with durations of a few picoseconds and very high-repetition rates. In principle, arbitrarily complex optical waveforms can be synthesised at high repetition rates by careful phase-intensity spectral shaping of frequency comb sources [1]. It remains however of high importance to have simple experimental methods for optical waveform generation as well as a clear mathematical description of the generated pulse profiles. Sinusoidal intensity profiles can be efficiently generated by use of standard high-bandwidth modulators. Gaussian and hyperbolic secant pulses are routinely delivered from fibre lasers. Within the toolbox of optical signal processing, triangular, parabolic and rectangular shapes can be achieved by linear [2] or nonlinear [3, 4] sculpturing. In the context of linear shaping or nonlinear fibre propagation, the properties of other pulse waveforms have also been the subject of recent discussion, including solitons over finite background [4, 5], superregular breathers [6], Riemann waves [7], flaticon waves [8], Airy pulses [9], and Hermite-Gaussian structures [10]. Some of these waves feature very strong oscillations in their temporal profiles [8-10].

An attractive method to generate high-repetition-rate and stable pulse trains relies on the spectral processing of periodically phase-modulated continuous-wave light [11]. Within this approach, the use of a quadratic spectral phase profile has indeed enabled the generation of ultra-short pulses at high repetition rates [12, 13] as well as of flat-top profiles [14]. We have recently introduced the line-by-line application of π/2 spectral phase shifts to a periodically phase-modulated continuous wave as an enhanced technique to obtain high-quality ultra-short pulse trains [15]. This pulse generation method can sustain multi-wavelength modulation and temporal multiplexing [16], as well as operation with a dual-tone signal [17]. Furthermore, compared to the original approach, the extinction ratio of the pulses and the suppression of sidelobes in their temporal intensity profiles are remarkably enhanced [15]. However, to date, no clear analytical description and explanation of the generated pulses have been reported. The discussion has essentially relied on numerical simulations with the amplitude of the initial phase modulation restricted to 2 radians.

In this paper, we fill this gap by providing an insight into the new type of an optical wave structure generated with this method, which we name 'besselon'. We unveil its properties for a broad range of phase-modulation amplitudes, and we derive simple analytical guidelines to predict the main pulse features. An approximate model is also introduced to help explain the evolution of the wave profile. Further, we theoretically discuss the possibility of doubling the pulse repetition rate, and identify optimum operating conditions to realise this. Our analytical predictions are validated by the experimental demonstration of the generation of besselon pulse patterns at repetition rates of 14 GHz and 28 GHz.

## II. Theoretical model for besselon waves

### A Pulse generation and properties

We consider a continuous optical wave with amplitude $\psi_0$ and carrier angular frequency $\omega_c$, $\Psi(t) = \psi_0 \, \psi(t) \, e^{i \omega_c t}$ whose phase is temporally modulated by a sinusoidal wave:

$$\psi(t) = e^{i \, A_m \cos(\omega_m t)}$$

where $A_m$ is the amplitude of the phase modulation and $\omega_m$ is its frequency. As a result of this phase modulation, the spectrum of frequency components of the wave envelope will consist of a series of discrete lines evenly spaced by $\omega_m$, which can be obtained from the Jacobi-Anger expansion [18-20] of $\psi(t)$:

$$\psi(t) = \sum_{n=-\infty}^{\infty} i^n \, J_n(A_m) \, e^{i n \omega_m t}$$

Therefore, the $n^{\text{th}}$ spectral component will have an intensity proportional to $J_n^2(A_m)$, where $J_n(x)$ is the Bessel function of the first kind and order $n$. An important characteristic of this spectrum is the existence of a $\pi/2$ phase shift between successive frequency components. The application of a quadratic spectral phase profile - such as produced by propagation through a purely dispersive element - to $\psi(t)$, with a curvature chosen so as to optimise the peak power of the resulting pulses, will enable partial compensation of the initial sinusoidal phase. This will in turn lead to the emergence of temporally localised, periodic pulse structures with a period $T_0 = 2\pi / \omega_m$ [12, 13], as shown in Fig. 1(b). We can see in Fig. 1(b) that increasing the amplitude $A_m$ of the initial modulation causes shortening of the central part of the pulses accompanied by a concomitant increase of the pulse peak power. However, since the initial spectral phase is not perfectly cancelled, the presence of a residual background impairs the resulting pulse intensity profiles [14]. On the other hand, the recent advances in linear pulse shaping technology make it now possible to manipulate optical frequency combs line by line. With our pulse-shaping approach, imprinting a π/2 phase shift to each individual spectral component of the phase-modulated continuous wave enables the synthesis of a new optical pulse field, which we call 'besselon' because of the important role played by Bessel functions in the description of the waveform features. The besselon wave envelope, $\psi_B$, can be represented as:

$$\psi_B(t) = \sum_{n=-\infty}^{\infty} J_n(A_m)\, e^{i n \omega_m t} = J_0(A_m) + 2\sum_{n=1}^{\infty} J_n(A_m)\cos\left(n\, \omega_m\, t\right) \qquad (1)$$

In Ref. [15], we have emphasised that this pulse structure exhibits an excellent extinction ratio (defined as the ratio of the power at $t = T_0/2$ to the power at $t = 0$) combined with the absence of spurious pedestals for values of the initial phase-modulation amplitude $A_m$ around 1.1 rad. Figure 1(a) shows the evolution of the wave intensity profile $|\psi_B|^2$ with $A_m$ across a wide range of modulation amplitude values, and features distinctly different from that of the pulse pattern resulting from spectral manipulation with a quadratic spectral phase. The besselon pulses do not experience a decrease of their temporal duration with increasing modulation amplitude, and their extinction ratio may also vary significantly with varying amplitude. The central part of the pulses also evolves in a different manner: sidelobes develop and tend to merge with the central part that displays an increasingly large number of strong oscillations.

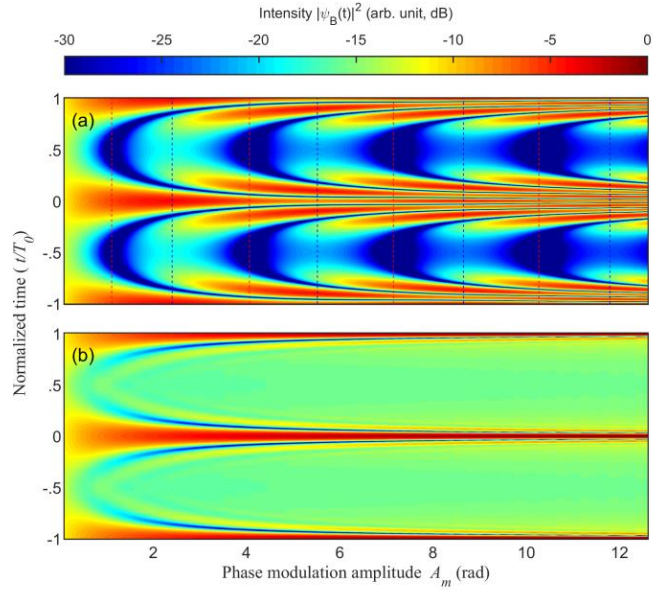

**Figure 1:** Evolution of the wave intensity profile with the amplitude of the initial sinusoidal phase modulation. **(a)** Besselon intensity profile $|\psi_B|^2$ as given by Eq. (1), and **(b)** profile synthesised by use of an optimal quadratic spectral phase profile as obtained from numerical simulations. The wave profiles are displayed using a logarithmic scale color map. The red vertical dashed lines in panel (a) represent the values of $A_m$ satisfying the condition for maximal extinction ratio, $I_{J0}(A_m) = 1$. The blue vertical dashed lines represent the zeros $A_{m,p}$ of $J_0(A_m)$, corresponding to the locations of the maxima of the peak power of the besselon.

To gain better insight into the besselon pulse wave, we have plotted on Fig. 2 the evolution of the pulse peak power at $t = 0$, the minimum power at $t = T_0/2$ and the extinction ratio of the pulse train with the phase-modulation amplitude. The Fourier-transform limited nature of the besselon wave for $A_m \leq 2.40$ rad brings about a continuous increase of the peak power with $A_m$, which reaches higher values than the peak power of the pulse wave obtained after a dispersive element. For further increase of the modulation amplitude, however, the peak power of the besselon wave oscillates and remains below the value of 6.10 obtained at $A_m = 2.40$ rad. From Eq. (1), one may derive analytically the value of the maximum of the field (i.e. at $t = 0$) :

$$\psi_B(t=0) = J_0(A_m) + 2\sum_{n=1}^{\infty} J_n(A_m)$$

and using the relations [18, 21]

$$2\sum_{n=1}^{\infty} J_{2n}(A_m) = 1 - J_0(A_m)$$

$$2\sum_{n=0}^{\infty} J_{2n+1}(A_m) = \int_0^{A_m} J_0(x)\, dx$$

we can obtain the peak amplitude of the besselon:

$$\psi_B(t=0) = 1 + I_{J0}(A_m) \tag{2}$$

with $I_{J0}$ being plotted on Fig. 2(d) (red line) and defined as

$$I_{J0}(A_m) = \int_0^{A_m} J_0(x)\, dx \tag{3}$$

We can see in Fig. 2(a) that the peak-power predictions from Eq. (2) are in perfect agreement with the results of numerical simulations. The amplitude values $A_{m,p}$ of the phase modulation at which the peak power of the besselon is maximal are given by $dI_{J0}/dA_m|_{A_{m,p}} = 0$, that is $J_0(A_{m,p}) = 0$. In other words, the maxima of the peak power occur at the zeros of the Bessel function of zero order. For these amplitudes ($A_{m,p}$ = 2.4048, 5.5201, 8.6537, …), the central frequency component of the spectrum vanishes and the pulse train becomes carrier-suppressed. Using the following asymptotic form of the zeroth-order Bessel function for large arguments [18] (Fig. 2(d), black circles)

$$J_0(x) \simeq \sqrt{\frac{2}{\pi x}} \cos\left(x - \frac{\pi}{4}\right) \tag{4}$$

we can obtain an approximate expression for the location of the $p^{\text{th}}$ maximum:

$$A_{m,p} \simeq \frac{3\pi}{4} + (p-1)\pi, \quad p \geq 1 \tag{5}$$

Equation (5) indicates that the power at $t=0$ of the besselon experiences periodic growth and decay with a period of $\pi$ rad. We also note that as $I_{J0}(x)$ asymptotically tends to 1 for large values of $x$ [21], the amplitude of the oscillations decreases with increasing values of $A_m$.

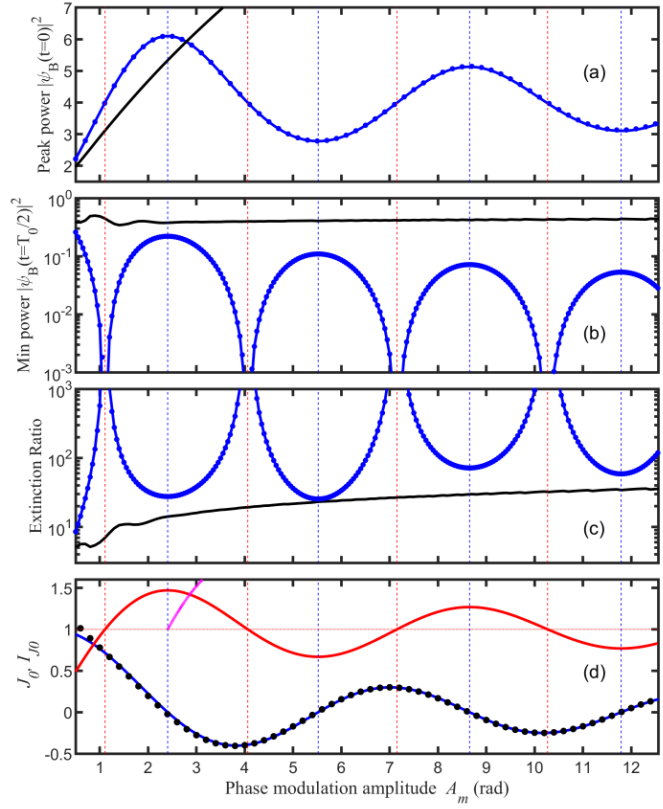

**Figure 2: (a-c)** Evolution of the **(a)** peak power, **(b)** minimum power, and **(c)** extinction ratio of the besselon pulse wave with the amplitude of the initial sinusoidal phase modulation. The predictions provided by (2), (6) and (7) (blue circles) are compared with the results of numerical simulations (blue curves). Also shown are the numerical simulation results for the pulse wave synthesised by use of an optimal quadratic spectral phase profile (black curves). **(d)** Bessel function of the first kind and order 0, $J_0(A_m)$, (blue curve) and its integral $I_{J0}(A_m)$ as defined by Eq. (3) (red curve). The filled black circles represent the asymptotic form of $J_0(A_m)$ for large argument (Eq. (4)). The red vertical dashed lines represent the values of $A_m$ satisfying the condition for maximal extinction ratio, $I_{J0}(A_m) = 1$. The blue vertical dashed lines represent the zeros $A_{m,p}$ of $J_0(A_m)$, corresponding to the locations of the maxima of the peak power. The purple curve in panel (d) represents the function $1 - 2 J_0(A_m)$ (Eq. (19) in section D).

The minimum power of the besselon pulse wave (Fig. 2(b)) also displays strong variations with varying amplitude of the initial phase modulation. This contrasts with the pulse wave obtained after a dispersive element, where the intensity level between consecutive pulses remains almost constant for $A_m$ above 2 rad. Similarly to the temporal maximum of the besselon, we can obtain the minimum amplitude of the besselon from Eq. (1):

$$\psi_B\left(t = \frac{T_0}{2}\right) = 1 - I_{J0}(A_m), \tag{6}$$

which yields the following expression for the extinction ratio of the pulses:

$$ER(A_m) = \frac{1+I_{J0}(A_m)}{1-I_{J0}(A_m)} \qquad (7)$$

We can therefore deduce that for amplitudes of the sinusoidal modulation satisfying the equation $I_{J0}(A_m) = 1$, the background level between pulses is zero hence the extinction ratio is maximal. This condition is consistent with the condition derived in [15] based on numerical simulation data. Similarly to the peak and minimum powers, the extinction ratio of the pulses varies strongly with $A_m$, while remaining larger than the extinction ratio of the pulse wave synthesised by use of a quadratic spectral phase profile across the whole $A_m$ variation range. Note that, for $I_{J0}(A_m) = 1$, the maximum amplitude of wave is simply 2, leading to an intensity of 4.

## B Details of the pulse shape

We now try to better describe the shape of the besselon at the points of maximal extinction ratio, i.e. when $I_{J0}(A_m) = 1$. Temporal profiles $\psi_B$ obtained for $A_m$ = 1.1086, 4.0628, 7.15 and 10.2695 rad are plotted on Fig. 3 (panels a). From Eq. (1), we can first note that the besselon is a waveform without any complex part, contrary to the waveform achieved after quadratic phase compensation. If the besselon does not present any negative part for $A_m < A_{m,1}$, this is not the case for higher values of $A_m$ and we can clearly see negative and positive parts that alternate, stressing that two successive oscillations of the wave are $\pi$-shifted. We also note that for an amplitude of oscillation $A_{m,p}$, we obtain *2p+1* peaks in the temporal intensity profile.

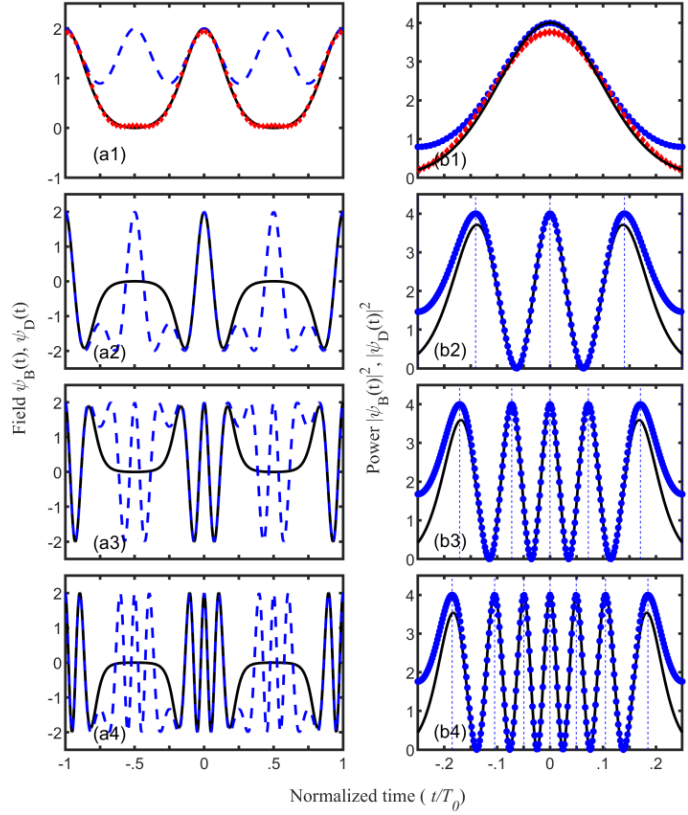

**Figure 3:** Temporal **(a)** amplitude and (b) power of the besselon for various values of $A_m$ leading to an optimum extinction ratio (the condition $I_{J0}(A_m) = 1$ is fulfilled). Results obtained for $A_m = 1.1086, 4.0628, 7.15$ and $10.2695$ are plotted on panel 1, 2, 3 and 4 respectively. The profile of the besselon $\psi_B$ (solid black line, Eq. (1)) is compared with the profile of $2\psi_D$ (dashed blue line, Eq. (8)). Approximation of the besselon proposed by Eq. (12) is plotted with red diamonds. The vertical dashed line in panels (b) indicate the position of the peak as predicted by Eq. (9).

In order to better understand the origin and the properties of these peaks, it could be interesting to express the besselon as:

$$\psi_B(t) = J_0(A_m) + 2\sum_{n=1}^{\infty} J_{2n}(A_m) \cos(2n\,\omega_m\,t) + 2\sum_{n=0}^{\infty} J_{2n+1}(A_m) \cos((2n+1)\,\omega_m\,t)$$

and to take into account that [18, 21]:

$$J_0(A_m) + 2\sum_{n=1}^{\infty} J_{2n}(A_m) \cos(2n\,\omega_m\,t) = \cos(A_m \sin(\omega_m\,t)) = \psi_D(t) \qquad (8)$$

It is interesting to superimpose the profiles of 2 $\psi_D$ and $\psi_B$ (see Fig. 3, blue dashed line and solid black lines respectively) and to note that the central part of the besselon is remarkably well described by 2 $\psi_D$. One may also note that the repetition rate of $\psi_D$ is twice the repetition rate of the besselon and that 2 $\psi_D(t) = \psi_B(t) + \psi_B(t-T_0/2)$. Note that $\psi_D$ can be generated directly from a Mach-Zehnder device driven by a sinusoidal electrical wave. From 2 $\psi_D$, it is possible to deduce several interesting features of the central part of the besselon. First the power of the various peaks equals 4 (amplitude of 2). Moreover, the temporal location of the maxima $t_M$ of the temporal intensity profiles are:

$$t_M = \operatorname{asin}\left(\frac{n\pi}{A_m}\right)\bigg/\omega_m \qquad (9),$$

with $n \leq (N-1)/2$, $N$ being the number of peaks that is given by $\lfloor 2A_m/\pi \rfloor + 1$. We can also derive the full width at half maximum of the besselon:

$$t_B = \frac{2}{\omega_m}\operatorname{asin}\left(\frac{\pi}{A_m}\left(\frac{N}{2}-\frac{1}{4}\right)\right) \qquad (10),$$

as well as the temporal width of the central part:

$$t_C = \frac{2}{\omega_m}\operatorname{asin}\left(\frac{\pi}{4A_m}\right) \qquad (11).$$

Results of those analytical predictions are reported on Fig. 4 where we can make out an excellent agreement with the numerical simulations. The difference between the waveform resulting from a quadratic spectral phase compensation and the besselon is also highlighted.

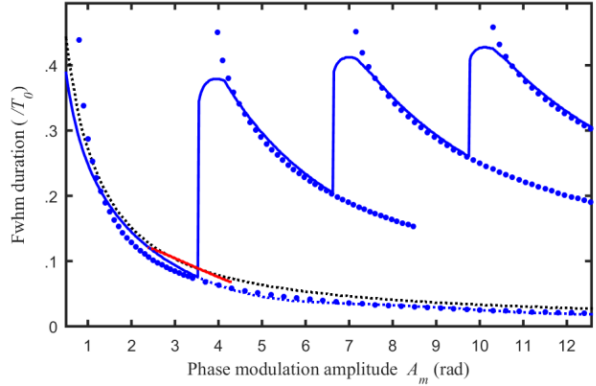

**Figure 4:** Evolution of the full-width at half maximum duration of the besselon wave. Numerical simulations (blue solid line) are compared with the analytical predictions of Eq. (10) and (11) (blue circles). The fwhm duration of the pulse obtained after optimal quadratic spectral phase compensation is plotted with black dotted line, whereas the duration of the central part of the besselon is plotted with dotted blue line. Results obtained for $\psi_B'$ are plotted with red line.

## C Simplified expression for low values of $A_m$

In this section, we now focus on a simplified model that efficiently describes the central part of the pulse for low values of $A_m$. Indeed, from Fig. 3(a1) and (b1) it appears that $2\psi_D$ is not accurate enough for low values of $A_m$ (typically below 1.5 rad). In this context, we have shown in previous works that for $A_m = 1.1$ rad, the generated waveform was very close from a Gaussian profile [15]. We have also recently discussed the similarity between the besselon profile obtained for low $A_m$ and the typical profile of an Akhmediev breather [22] that may exist in nonlinear fibre optics [23]. Here, we propose to interpret the waveform from another viewpoint. As shown in Fig. 5(a), we can consider that the spectrum obtained for low $A_m$ is the result of the coherent superposition of three waves: a continuous wave $\psi_1$ with an amplitude $J_0-2J_2$ and two sinusoidally partially modulated wave (a continuous background with an amplitude $J_1$ that is modulated with an amplitude $2 J_2$ at a frequency $\omega_m$) that are frequency shifted by $-\omega_m$ and $\omega_m$ ($\psi_2$ and $\psi_3$ respectively) In the temporal domain, this leads to

$$\psi_B(t) \simeq \left[J_0 - 2J_2\right] \\ + \left[\left(J_1 + 2J_2 \cos(\omega_m t)\right) e^{i\omega_m t}\right], \\ + \left[\left(J_1 + 2J_2 \cos(\omega_m t)\right) e^{-i\omega_m t}\right]$$

which can be rewritten as

$$\psi_B(t) \simeq [J_0 - 2J_2] + 2[J_1 + 2J_2 \cos(\omega_m t)]\cos(\omega_m t) \quad (12),$$

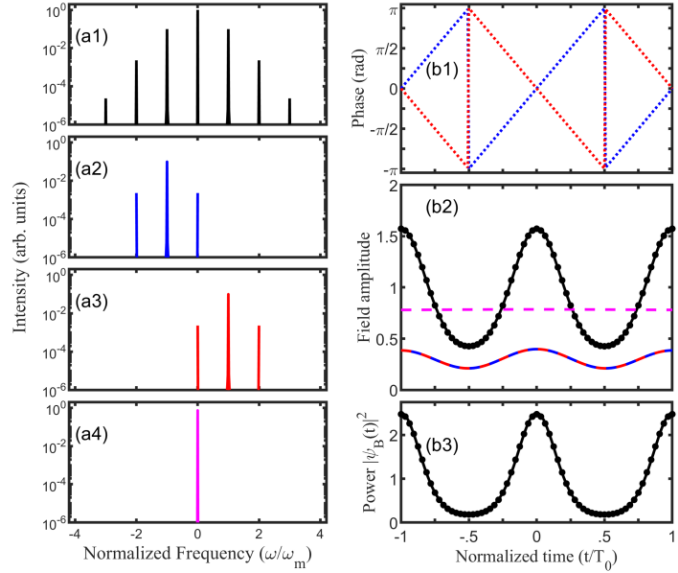

**Figure 5:** Comparison of the besselon properties obtained for $A_m = .6$ rad with the predictions of a simplified 3-wave model. **(a)** Optical spectrum of $\psi_B$, $\psi_2$, $\psi_3$ and $\psi_1$ (panels 1,2,3,4 respectively). **(b)** Temporal properties: chirp profile, field and intensity profiles (panel 1, 2, 3, respectively). The results for waves $\psi_1$, $\psi_2$ and $\psi_3$ are plotted with purple, blue and red colors respectively. Results $\psi_B$ are plotted with solid black line and compared with the approximate model (black circles, Eq. (12)).

The amplitude and phase profiles of the three waves are plotted in Fig. 3(b1-b2) which helps to understand how the waveform is formed. For $t = 0$, the three waves are in phase, they interfere constructively and the amplitude is maximum so that

$$\psi_B(t=0) \simeq J_0(A_m) + 2J_1(A_m) + 2J_2(A_m) \quad (13),$$

On the contrary, for $t = T_0/2$, we note that the modulated waves are both $\pi$-shifted with respect to the continuous wave. They will interfere destructively and the amplitude of the besselon is minimum as follows:

$$\psi_B(t=T_0/2) \simeq J_0(A_m) - 2J_1(A_m) + 2J_2(A_m) \quad (14),$$

The besselon therefore reaches a null value at $T_0/2$ when

$$J_0(A_m) = 2(J_1(A_m) - J_2(A_m)) \qquad (15),$$

that gives an optimum value of $A_m = 1.187$ rad, which is consistent with the value found previously (1.109 rad). Finally, we can note that for $t = T_0/4$, the two sinusoidal waves present a phase difference of $\pi$. Consequently, they cancel each other and $\psi_B(t=T_0/4) \simeq J_0(A_m) - 2J_2(A_m)$. Therefore, for low values of $A_m$, $\psi_B(t=T_0/4)$ continuously decreases indicating that the pulse gets temporally shorter and shorter. We have plotted the intensity profiles resulting from the approximate Eq. (12) on Fig. 5(b3) for a value of $A_m = .6$ rad. The approximation is nearly undiscernible compared to the exact prediction from Eq. (1). The agreement is also rather good for $A_m = 1.10$ rad as can be seen on Fig. 3(a1-b1). In Fig. 6(a-b), we compare the peak-power and extinction ratios predicted by this approximate model (purple circles) and once again, we find an excellent agreement as long as the level of the component $J_3$ is low enough.

## D Reduction of the lateral temporal sidelobes

As can be seen in Fig. 1(a) and in the magnified view provided in Fig. 6(c1), the evolution is marked for $A_m > A_{m,1}$ by the development of noticeable sidelobes in the pulse. Given the non-negligible energy fraction contained in those sidelobes, this leads to a decrease of the peak power as stressed in panel (a) of Fig. 6. Recall that the value $A_{m,1}$ corresponds to the amplitude of the initial phase modulation for which the central spectral component decreases down to zero. It then becomes negative up to to $A_m = A_{m,2}$. Consequently, from the simplified model we developed in the previous section, we can understand that at $t = 0$ and between $A_{m,1}$ and $A_{m,2}$, the continuous background becomes $\pi$-shifted with respect to the sinusoidally modulated waves. Therefore, the interference changes from constructive interference (below $A_m < A_{m,1}$) to destructive interference. The resulting structure has a reduced peak power and is not Fourier-transform limited anymore. In order to overcome this effect, a solution is to imprint an additional $\pi$ shift on the central component for $A_m > A_{m,1}$. In this case, the modified besselon pulse $\psi'_B$ is provided by $\psi'_B(t) = \psi_B(t)$ for $A_m < A_{m,1}$ and:

$$\psi'_B(t) = -J_0(A_m) + 2\sum_{n=1}^{\infty} J_n(A_m) \cos(n\omega_m t) = \psi_B(t) - 2J_0(A_m) \qquad (16)$$

elsewhere. The corresponding peak amplitude becomes for $A_m > A_{m,0}$:

$$\psi'_B(t=0) = 1 + I_{J0}(A_m) - 2J_0(A_m) \qquad (17)$$

The maximum amplitude is therefore increased by $2|J_0(A_m)|$ with respect to the case without $\pi$ phase shift. The condition on $A_m$ that now leads to a maximum amplitude is that $J_0(A_m) = -2J_1(A_m)$, leading to a value of the maximum for $A'_{m,1} = 3.40$ rad (cyan vertical line in Fig. 6a) for a peak-power value that is 8.93, representing a 46% increase compared to the results obtained for $\psi_B$. The extinction ratio becomes:

$$ER'(A_m) = \frac{1 + I_{J0}(A_m) - 2J_0(A_m)}{1 - I_{J0}(A_m) - 2J_0(A_m)}, \qquad (18)$$

which tends to infinity when:

$$I_{J0}(A_m) = 1 - 2J_0(A_m) \qquad (19)$$

which leads to a condition $A_m = 2.86$ rad as can been found from Fig. 1(d). This value differs quite significantly than the one found for the unmodified besselon. Note that an ER above $10^2$ is obtained for $A_m$ in the range of 2.6-3.15 rad. The analytical results from Eq. (17), (18) are in perfect agreement with the numerical results as can be seen on Fig. 6(a-b). Comparison of the temporal profiles obtained without or with an additional $\pi$ phase shift on the central component is provided in Fig. 6(c). It clearly highlights the decrease of the sidelobes of the pulse that are at -13 dB of the peak power for $A_m = 2.86$ rad (while in the case of $\psi_B$ the decrease is only -7.22 dB for the same $A_m$). If there is an improvement in terms of peak-power, we do not observe such an improvement in terms of temporal duration (see Fig. 4). For $A_m = 2.86$ rad, pulse with a fwhm temporal duration of 0.1 $T_0$ is obtained, which represents a significant reduction when compared to the results achieved for $A_m = 1.1086$. Finally, note that for $A_m > 3.83$, the first lateral sidebands will experience a $\pi$ phase shift as $J_1$ changes of sign. Therefore, insertion of additional phase shifts could be required to further improve the peak-power.

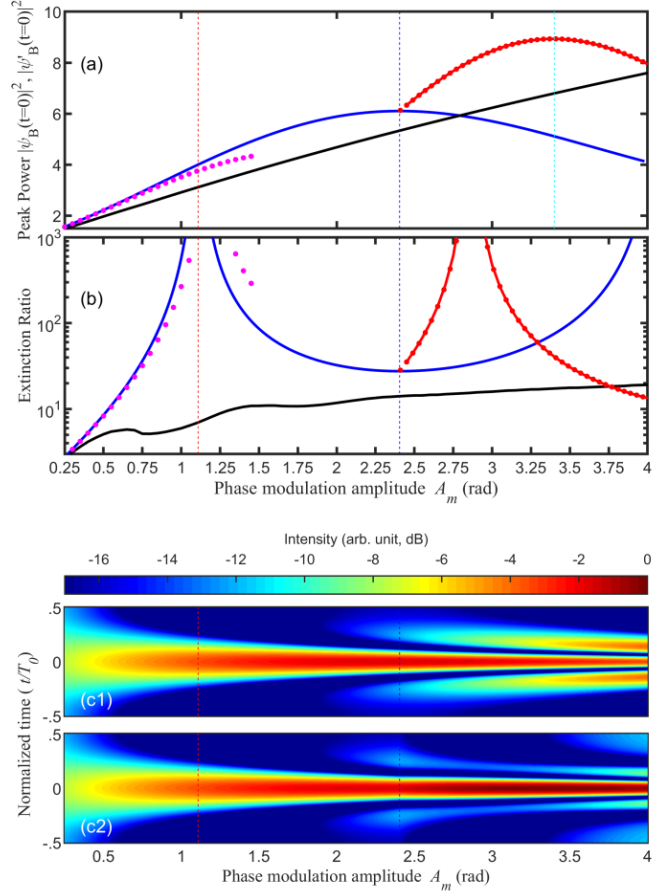

**Figure 6: (a-b)** Evolution of the properties of the temporal intensity profile according to the initial amplitude of the sinusoidal phase modulation. **(a)** Peak power. **(b)** Extinction ratio. Results obtained after optimum quadratic phase compensation (black solid lines) are compared with the properties of the besselon (blue lines, Eq. (2) and (7)) and of the modified besselon (red lines, analytical predictions solid line, analytical predictions – Eq. (17) and (18), red circles). Results from the approximate model (derived from Eq. (12)) are plotted with purple circles. **(c)** Intensity profiles obtained for the besselon $|\psi_B|^2$ (Eq. (1)) and modified besselon $|\psi'_B|^2$ (Eq. (16)).

## E Time multiplexing

In this last section, we consider the possibility to temporally interleave the besselon pulses $\psi'_B$ in order to double the repetition rate. Many solutions may exist such as the use of the fractional Talbot effect [24, 25], temporal delay lines [26], or spectrum manipulation such as intensity modification [27] or phase only shaping [28]. Here, in order to double the repetition rate, we suppress all odd spectral components. In this context, the resulting field is restricted to:

$$\begin{cases} \psi'_{2B}(t) = J_0(A_m) + 2\sum_{n=1}^{\infty} J_{2n}(A_m) \cos(2n\,\omega_m\,t) & \text{for } A_m \leq A_{m,1} \\ \psi'_{2B}(t) = -J_0(A_m) + 2\sum_{n=1}^{\infty} J_{2n}(A_m) \cos(2n\,\omega_m\,t) & \text{for } A_m > A_{m,1} \end{cases}$$

That, given Eq. (8), can be rigorously expressed as :

$$\begin{cases} \psi'_{2B}(t) = \psi_D(t) & \text{for } A_m \leq A_{m,1} \\ \psi'_{2B}(t) = \psi_D(t) - 2\,J_0(A_m) & \text{for } A_m > A_{m,1} \end{cases} \tag{20}$$

Results of the repetition doubled pulse train are shown on Fig. 7(a). One can make out that the optimal condition in terms of extinction ratio is obtained for $A_m = 1.57$ and $A_m = 3.78$ and not for $A_m = 1.10$ and $A_m = 2.86$ rad as one could have initially anticipated from sections IIA,D (Eq. (7) and Eq. (19)). Indeed, the value at $t = T_0/4$ is given by:

$$\begin{cases} \psi'_{2B}(t = T_0/4) = \cos(A_m) & \text{for } A_m \leq A_{m,1} \\ \psi'_{2B}(t = T_0/4) = \cos(A_m) - 2\,J_0(A_m) & \text{for } A_m > A_{m,1} \end{cases}$$

and vanishes when

$$\begin{cases} A_m = \pi/2 & \text{for } A_m \leq A_{m,1} \\ 2\,J_0(A_m) = \cos(A_m) & \text{for } A_m > A_{m,1} \end{cases} \tag{21}$$

In Fig. 7(b), we compare the profiles obtained before repetition rate doubling (panel 1) with the results achieved after processing (panel 2). We can make out that suppressing the odd components enables to get rid of the residual component observed between two pulses [26]. Nice pulse shapes are achieved especially for $A_m = 3.78$ rad (optimum operating point for $A_m > A_{m,1}$) where the spurious sidelobes are decreased down to a level of nearly -20dB the peak power. Temporal duration of 0.092 $T_0$ leads to a duty cycle of 0.18. The cost of the approach is mainly in the drop of the peak power, which decreases by a factor more than two.

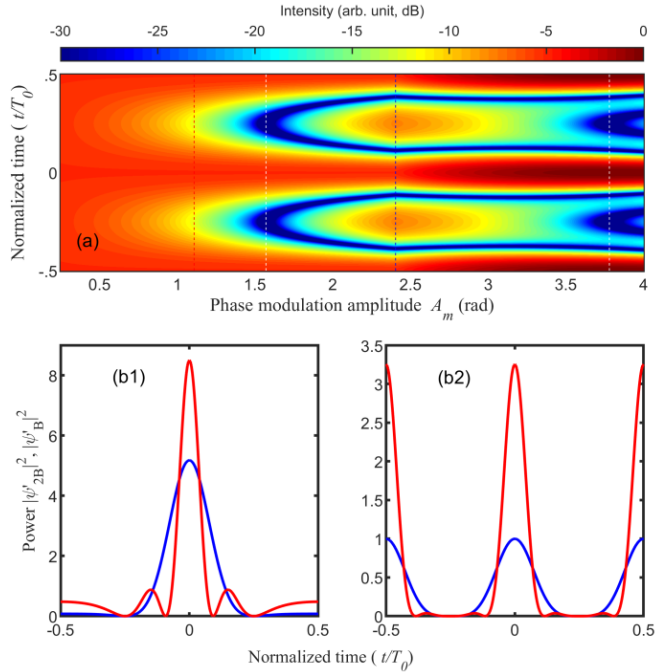

**Figure 7 :** **(a)** Evolution of the intensity profile obtained for modified besselon $|\psi'_{2B}|^2$ at the doubled repetition rate (Eq. **Erreur ! Source du renvoi introuvable.**). (b) Detail of the temporal intensity profile obtained before and after repetition rate doubling $\psi'_B|^2$ and $|\psi'_{2B}|^2$ (panels 1 and 2 respectively), for $A_m = \pi/2$ and 2.78 rad respectively (blue and red curves).

# III. Experimental validation

## A. Experimental setup

The experimental setup is sketched in Fig. 8(a) and is based on devices that are commercially available and typical of the telecommunication industry. A continuous wave laser at 1550 nm is first temporally phase modulated using a Lithium Niobate electro-optic device driven by a sinusoidal electrical signal at a frequency $f_m = 14$ GHz. The use of a low $V_\pi$ modulator enables us to investigate modulation amplitudes $A_m$ up to 3.75 rad. Note that amplitudes of modulations exceeding 10 rad have already been demonstrated in the past using a resonant microwave modulator [29, 30] or cross-phase modulation in a highly non-linear fiber [8].

A linear spectral shaper (Finisar Waveshaper) based on liquid crystal on silicon technology [31] is then used to apply the suitable spectral phase profiles made of simple discrete spectral phase shifts of π/2 applied between two successive components for the experimental demonstration of the waves $\psi_B$ and $\psi'_B$ (see panels 8(b1) and 8(b2) respectively). For the doubling of the repetition rates, we have inserted π phase shifts at frequencies $(2p+1) f_m$, with $p \geq 0$ for the wave $\psi'_{2B}$ ($A_m \leq A_{m,1}$) and with $p \geq 1$ for ($A_m > A_{m,1}$) (see panels 8(b3) and 8(b4), red curves). Indeed, the discrete spectral π-phase shifts create a notch filter [32] that dramatically attenuates the spectral harmonics to be suppressed. Attenuation curves recorded using an amplified spontaneous emission source (ASE) as an input signal are shown with black solid lines and confirm that we can simultaneously achieve the spectral amplitude and phase targets using a simple phase-only spectral shaping. Let us mention that for operation at a fixed wavelength and a fixed repetition rate, the linear shaping stage can also be fully realized by a cascaded uniform fibre Bragg grating [33]. Note also that in order to increase the repetition rate by suppression one spectral component of two, other approaches can be implemented such as the use of a birefringent fibre [34] or Fabry-Perot interferometers [35].

The resulting signal is directly recorded by means of a high-speed optical sampling oscilloscope (1 ps resolution) and with a high-resolution optical spectrum analyzer. A low noise erbium-doped amplifier has been inserted in order to reach the optimum power level on the temporal detection device.

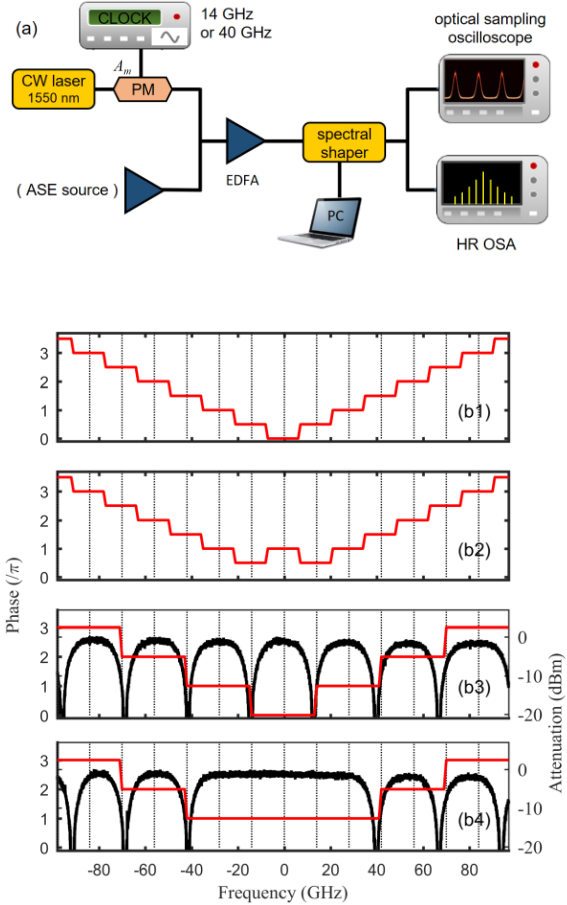

**Figure 8: (a)** Experimental setup : CW : Continuous Wave ; PM : Phase Modulator ; RF : Radio-Frequency ; ASE : Amplified Spontaneous Emission ; EDFA : Erbium Doped Fiber Amplifier ; HR OSA : High Resolution Optical Spectrum Analyzer. **(b)** Spectral phase profile (red line) applied on the programmable phase shaper for the generation of $\psi_B$, for $\psi'_B$ and for the simultaneous phase shaping and repetition rate doubling of $\psi_B$ and $\psi'_B$ (panels 1,2, 3 and 4 respectively). The optical attenuation induced by the phase attenuation is plotted with black line.

## B. Pulse waveform

A set of experimental temporal and spectral power profiles recorded for different levels of initial phase modulation $A_m$ is summarized on Fig. 9. The agreement between the experimental results and the theoretical predictions is excellent, both in the temporal and spectral domains. We can recognize all the various features that have been discussed in the preceding sections. For low values of $A_m$ (typically below 0.8), the wave is only partially modulated and a continuous background remains (panel a). For values around 1.1 rad (panel b), the extinction ratio becomes very high and a Gaussian-like waveform is obtained. For increasing values of $A_m$ (2.47 rad, panel c) the background impairs significantly the pulse train. Significant sidelobes then develop (panel d1). Those pedestals can be suppressed by inserting an additional π phase shift on the central spectral component (panel d3).

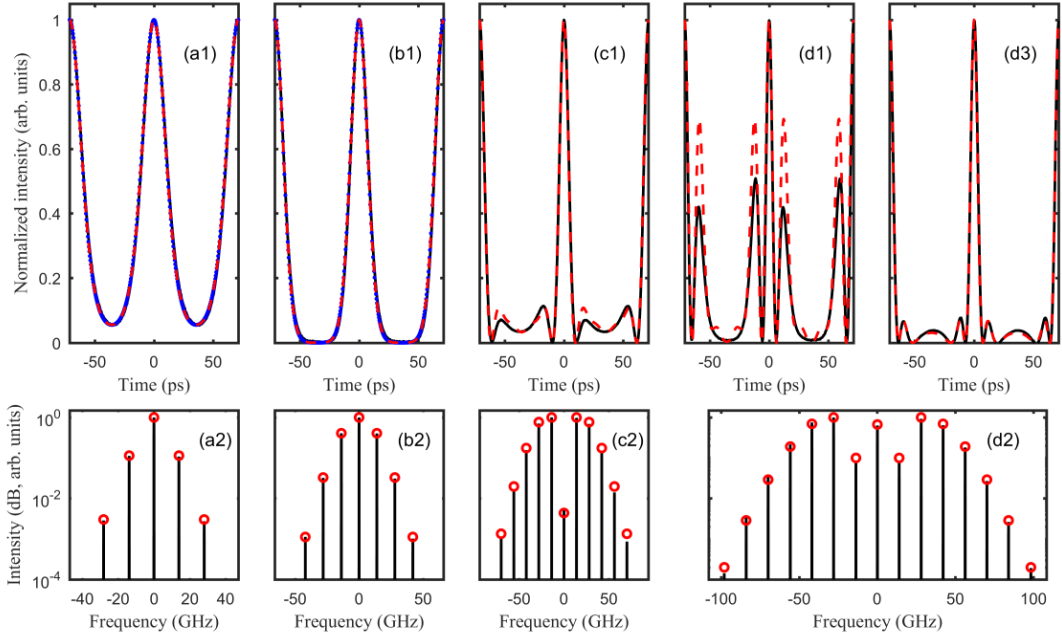

**Figure 9:** Temporal and spectral intensity profiles (panels 1 and 2, respectively) $\psi_B$ obtained for levels of amplitude modulation $A_m$ of 0.64 rad, 1.08 rad, 2.47 rad and 3.48 rad (panels a, b, c and d respectively). For 3.48 rad, the panel (d3) corresponds to the modified besselon $\psi'_B$. Experimental results (black lines) are compared with the theoretical predictions (red dashed lines and red circles, Eq. (1) or (16) for panel (d3)) as well as the predictions of the approximate model (blue dashed line, Eq. (12). All the profiles are normalized to their maximum.

The results of a more systematic experimental study are reported in Fig. 10 for both structure type $\psi_B$ and $\psi'_B$. These results are to be compared with the results obtained in Fig. 6(c) and we note once again a remarkable agreement, both in terms of central shape and also in terms of the evolution of the sidelobes. We have also put the details of the time profiles recorded at the compression point corresponding to the best extinction ratio (2.83 rad), which further stresses the beneficial impact of the π phase shift on the central spectral component.

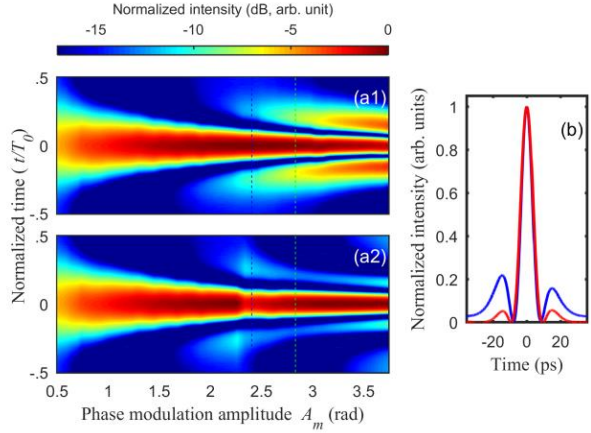

**Figure 10: (a)** Experimental evolution of the temporal intensity profile according to the initial amplitude of the sinusoidal phase modulation. Results for $\psi_B$ and $\psi'_B$ are plotted on panels (a1) and (a2) respectively. White dotted line corresponds to $A_m = 2.83$ rad whereas the black dotted line is for $A_m = A_{m,1}$. **(b)** Details of the intensity profile for $A_m = 2.83$ rad

The excellent agreement between theory and experiment is also visible in Fig. 11 where we have summarized the evolution of the peak-power and the fwhm duration. All the various features we previously predicted are reproduced, thus confirming our modelling.

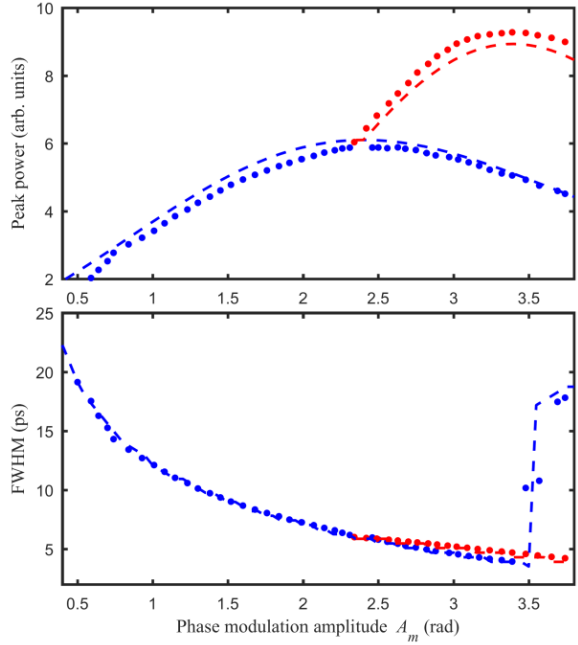

**Figure 11:** Evolution of (a) the peak-power and (b) the fwhm duration of the pulse train according to the amplitude of modulation $A_m$. The experimental results (circles) are compared with the results of Eq. (2) and (17) or with numerical simulations (dashed lines). Results obtained with the besselon $\psi_B$ and the modified besselon $\psi'_B$ are plotted with blue and red colors, respectively.

## C. Doubling of the repetition rate

We finally experimentally investigate the doubling of the repetition rate. Resulting pulse trains obtained at 28 GHz are reported in Fig. 12 for $A_m = \pi/2$ and $A_m = 3.72$ rad. The optical spectra are displayed in panels (2) where we can note the significant decrease of the odd spectral components. The $\pi$ phase shifts applied on those harmonics induces a reduction by more than 20 dB of their power. The odd spectral components are not fully suppressed but their very low level no longer influences the overall stability of the resulting train as it is confirmed by the eye diagrams. In both cases, the experimental results are in excellent agreement with the theoretical predictions form Eq. (20). The extinction ratio is excellent. The level of the pedestals is very low and for $A_m = 3.72$ rad, a fwhm duration of 6.9 ps is achieved, leading to a duty-cycle of 0.19.

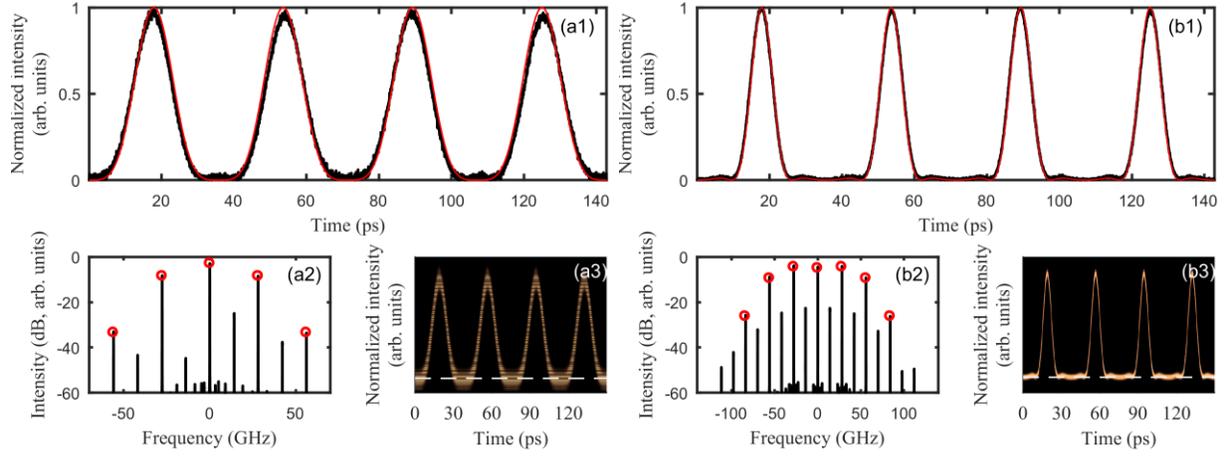

**Figure 12:** Pulse trains generated at a repetition rate of 28 GHz obtained for (a) $A_m = \pi/2$ rad and (b) $A_m = 3.72$ rad. (panel 1) Temporal power profile. (panel 2) Optical power spectrum. (panel 3) Temporal profile recorded in persistent mode. Experimental results (black line) are compared with theoretical results (red line, Eq. (20); circles). The horizontal white dashed lines indicate the zero-level.

# IV. Conclusions

In conclusion, we have theoretically introduced a new waveform we have named besselon for which we have provided accurate and simple analytical predictions of the main features. We have stressed the significant differences between the besselon and the waveform that arises from quadratic phase compensation of a sinusoidally phase modulated wave. For a low level of initial phase modulation, an approximate model has been proposed to interpret the besselon shape as the result of an interference between three waves. For high levels of initial phase modulation, the single pulse structure then splits into several peaks. By inserting an additional $\pi$-phase shift on the spectral component, the peak power can be further increased and the temporal duration reduced, making it attractive for high-repetition sources as well as for temporal multiplexing. Duty cycles as low as 0.1 without residual background are demonstrated without involving another complex stage of fiber-based nonlinear compression [36, 37]. By carefully choosing the level of initial modulation, we have shown that efficient repetition-rate doubling can be achieved with nice temporal profiles free from residual background. Compared to existing solutions that have relied on the use of an additional intensity modulator tightly synchronized with the phase modulation [38, 39] or on the use of a nonlinear optical loop mirror [40], our proposed method only implies a single phase modulator. Our approach could also be of interest for noise-free amplification [41] as well as for optical sampling [42]. In the present contribution, we have mainly focused on the generation of the pulse waveform. Further studies will explore the propagation in linear media (where it will be subject to Talbot self-imaging [24]) or in nonlinear media such as anomalous dispersive fibers where the spectrum with many components could be of extreme interest to seed new nonlinear regimes.

## Acknowledgements

We thank Julien Fatome for stimulating initial discussions. We acknowledge the support of the Institut Universitaire de France (IUF), the Bourgogne-Franche Comté Region, the French Investissements d'Avenir program and the Agence Nationale de la Recherche (ISITE-BFC ANR-15-IDEX-0003 Project Bright). Experiments have benefited from the PICASSO experimental platform of the University of Burgundy.